\documentstyle[11pt,paspconf,twoside]{article}  
\input psfig.sty

\def\NH{$N_{\rm HI}$}  
\def \hi {H\,{\sc i~}}

\def\kms{km\,s$^{-1}$}

\begin{document}  
 
\title{Morphological Characteristics of Compact High--Velocity Clouds  
Revealed by High--Resolution WSRT Imaging}  
\author{W.B. Burton}  
\affil{Sterrewacht Leiden, P.O. Box 9513, 2300 RA Leiden, The Netherlands}  
\author{R. Braun}  
\affil{Netherlands Foundation for Research in Astronomy, P.O. Box 2, 7990  
AA Dwingeloo, The Netherlands}  
\begin{abstract} 
  A class of compact, isolated high--velocity clouds which plausibly
  represents a homogeneous subsample of the HVC phenomenon in a single
  physical state was objectively identified by Braun and Burton (1999).
  Six examples of the CHVCs, unresolved in single--dish data, have been
  imaged with the Westerbork Synthesis Radio Telescope. The
  high--resolution imaging reveals the morphology of these objects,
  including a core/halo distribution of fluxes, signatures of rotation
  indicating dark matter, and narrow linewidths constraining the
  kinetic temperature of several opaque cores.  In these regards, as
  well as in their kinematic and spatial deployment on the sky, the
  CHVC objects are evidently a dynamically cold ensemble of
  dark--matter--dominated \hi clouds accreting onto the Local Group in
  a continuing process of galactic evolution.
 
\end{abstract}  
 
 
\section{Introduction}  
The criteria applied by Braun and Burton (1999) to the  
Leiden/Dwingeloo \hi Survey of Hartmann \& Burton  
(1997) and the HVC catalog of Wakker \& van Woerden (1991)  
led to a catalog of 65 confirmed examples of compact, isolated  
high--velocity clouds.  The selection criteria excluded the Magellanic 
Stream and all other HVC  
complexes.  The catalog is more likely to represent a single  
phenomenon than would a sample which included the major HVC complexes.   
The CHVC objects plausibly originated under common circumstances,  
have shared a common evolutionary history, are arguably in a single  
physical state, and have not (yet)  
been strongly influenced by the radiation field of the Milky Way or  
of M31, or by a gravitational encounter with one of these major systems.  
In this context, the extended HVC complexes  
would be the nearby objects currently undergoing accretion onto the  
Galaxy, while the more compact, isolated ones would be their distant  
counterparts in the Local Group environment. The positional and  
kinematic characteristics of the compact HVCs are similar in many regards 
to  
those of the Local Group galaxies. The sample is distributed quite  
uniformly over the sky, and defines a  
well--organized kinematic system.  The kinematic signature of this  
system suggests an in--falling population associated with the  
Local Group gravitational potential.  
 
\begin{figure}[]  
\vspace{-7cm}  
\psfig{file=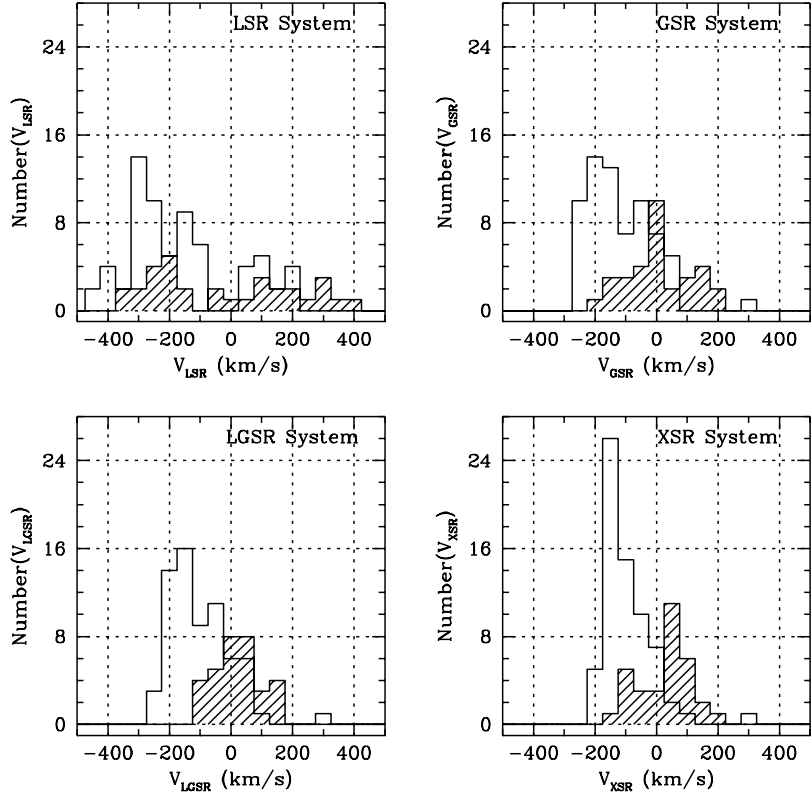,width=17cm}  
\vspace{-1.1cm}  
\vfill  
\caption{Histograms of distributions of the CHVC velocities, and of Local  
Group galaxy velocities, as measured in the indicated reference frames  
(Braun and Burton 1999).  The solid lines represent the CHVC  
ensemble; the shaded areas, Local Group galaxies.  The dispersion  
in the $v_{\rm LGSR}$ frame is  
significantly smaller than those in the $v_{\rm LSR}$ and $v_{\rm  
GSR}$ reference frames.  The velocities labelled $v_{\rm XSR}$  
refer to a frame which minimizes the dispersion of the  
CHVC velocity distribution; within the errors, this reference frame is  
consistent with the Local Group Standard of Rest. This minimization  
provides a quantitative demonstration of Local--Group deployment.  Within 
this  
frame, the CHVC ensemble is dynamically quite cold, with a dispersion of  
only 69~km~s$^{-1}$, although strongly in--falling into the Local Group  
barycenter at a velocity of about 100~km~s$^{-1}$.   
 } \label{fig:hist}  
\end{figure}  
 
\section{Minimized Velocity Dispersion of the CHVC Ensemble \\
as an Indication of Local Group Deployment}  
 
The possibility of the extragalactic nature of high--velocity clouds  
has been considered in various contexts by, among others, Oort  
(1966, 1970, 1981), Verschuur (1975), Eichler  
(1976), Einasto et al. (1976), Giovanelli  
(1981), Bajaja et al. (1987), Burton (1997),  
Wakker and van Woerden  
(1997), Braun and Burton  
(1999), and Blitz et al. (1999).  Nevertheless, no direct distance  
determination is yet available for any of the individual  
CHVC objects.  Distances are, of course, required to establish the values 
of important physical parameters: linear size varies as $D$, mass as  
$D^2$, and density as $D^{-1}$.   
 
Several aspects of the topology of the class are difficult  
to account for if the CHVCs are viewed as a Milky Way population, in  
particular if they are viewed as consequences of a galactic fountain.  
The amplitude of the horizontal motions of these ``bullets" is  
comparable to that of the vertical motions.  The vertical motions are  
larger than expected for free fall onto the Milky Way from material  
returning in a fountain flow.  The topology shows no preference for the  
terminal--velocity locus, where motions from violent events leading to  
a fountain would be expected to be most common.  Unlike the situation  
if the major HVC complexes are considered, the CHVCs are scattered  
rather uniformly across the sky. The CHVCs show no tendency to  
accumulate in the lower halo of the Milky Way.   
 
On the other hand, regarding both  
their spatial and kinematic distributions, the CHVCs show substantial  
similarities with the distributions of the galaxies comprising the  
Local Group.  
Since our CHVC sample has both a substantial size and a rather uniform 
distribution on the sky (see Fig.~1 of Braun and Burton 1999) it  
is appropriate to use the sample itself to define a best--fitting velocity  
reference system.  Braun and Burton (1999) showed that the velocity  
dispersion of the ensemble is minimized in a reference frame consistent  
with the Local Group Standard of Rest.  
The solar apex which follows directly from a minimization  
of the velocity dispersion of the CHVC system, namely,  
$(l_\odot,b_\odot,v_\odot) = (88^\circ,-19^\circ,+293$~km~s$^{-1}$),  
agrees within the errors with that which defines the Local Group  
Standard of Rest, $(l_\odot,b_\odot,v_\odot) =  
(93^\circ,-4^\circ,+316$~km~s$^{-1}$), found by Karachentsev \& Makarov  
(1996). The velocity dispersion of the CHVC system in this  
reference frame is only $\sigma_{\rm XSR}=69$~km~s$^{-1}$, while there  
is a mean in--fall of $v_{\rm LGSR}=v_{\rm XSR}=-100$~km~s$^{-1}$.  
 
Figure 1 shows histograms of the velocities in these  
reference frames.  The dispersion of the velocities decreases in a  
progression from the $v_{\rm LSR}$ reference frame, for which  
$\sigma_{\rm LSR} = 175$ km~s$^{-1}$, via the $v_{\rm GSR}$  
($\sigma_{\rm GSR} = 95$ km~s$^{-1}$) and the $v_{\rm LGSR}$ $\sigma_{\rm 
LGSR} = 88$ km~s$^{-1}$ frames, to the minimum of $\sigma_{\rm 
XSR}=69$~km~s$^{-1}$ for the $v_{\rm XSR}$ frame.  This minimization  
provides a quantitative demonstration of Local--Group deployment and, in  
addition, that the CHVC ensemble is dynamically quite cold.

\section{Imaging with the WSRT}  
 
The CHVC objects catalogued by Braun and Burton (1999) are not spatially  
resolved in the single--dish data of the Leiden/Dwingeloo Survey.  Insofar  
as the objects have not been resolved in angle, it has remained ambiguous  
if the single-dish linewidth refers to the intrinsic characteristics of a  
single entity or to the collective behavior of blended features.   
High--resolution imaging is required to reveal such kinematic properties  
as intrinsic linewidths and opacity information, or indications of rotational  
support requiring a higher total mass than available in the \hi alone, as  
well as the resolved structural morphology. If the CHVCs are in fact a  
population of primordial clouds scattered throughout the Local Group,  
then they might reveal some morphological  
characteristics which would not be consistent with the expectations  
of other suggested scenarios, in particular for  
objects ejected by a galactic fountain (e.g. Shapiro and Field 1976,  
Bregman 1980) or located within the Galactic  
halo (Moore et al. 1999).  High--resolution imaging is also necessary to  
provide specific targets for deep optical observations.  Such optical  
probes would help clarify the distinction between the CHVCs and  
(sub--)dwarf galaxies, and any indication of a stellar population would  
offer a distance indication.  
                           
Of the sample of 65 compact, isolated  
HVCs catalogued by Braun and Burton (1999), only two had been subject  
to interferometric imaging.  Wakker and Schwarz (1991) used the  
Westerbork array to show that both CHVC114$-$10$-$430 and  
CHVC111$-$06$-$466 are characterized by a core/halo morphology, with only  
about 40\% of the single--dish flux recovered on angular scales of tens  
of arcmin, and, furthermore, that the linewidths of the single--dish  
spectra of these two sources were resolved into components of some 5 \kms\,  
width or less.  
Both of the imaged systems display systematic velocity gradients along  
the major axis of an elliptical \hi distribution, which Wakker and  
Schwarz judged to be suggestive of rotation in self--gravitating systems  
at Local Group distances.  
 
Braun and Burton (2000) have used the Westerbork array to image an  
additional six objects of the CHVC class.  Although only six CHVC sources 
were imaged in our program, the sources are distributed widely in galactic 
coordinates, span radial velocities of $-275<v_{\rm LSR}<+165$ \kms, vary 
in single-dish linewidth from 6 to 95 \kms, and in line flux from 25 to 300 
Jy\,\kms.  One twelve--hour integration was obtained for each field in the 
standard WSRT array configuration, having a shortest baseline of 36 meters. 
The angular resolution retained for data presentation was about 
one arcmin, depending on the flux of the emission. The effective velocity 
resolution was 1.2 times the channel spacing of 2.06~km~s$^{-1}$.  Examples 
of the WSRT imaging discussed below reveal a characteristic core/halo 
arrangement of \hi fluxes, narrow linewidths, and, in several cases, a 
signature of rotation.

\section{Indications of a Core/Halo Morphology}  
 
     Moment images of the integrated \hi emission, together with several  
representative spectra, are shown in Figures 2, 3, 4, and 6 for four of the  
CHVCs observed with the WSRT.  In each case the emission is dominated by  
one or more bright knots, embedded in less intense, more diffuse gas.  The  
linewidths in the cores are narrower than in the halos.  
 
%\subsection{CHVC\,069+04$-$223}  
 
CHVC\,069+04$-$223, represented in Fig.~2, illustrates the characteristic  
morphology.  The \NH~ distribution is dominated by a bright  
elliptical concentration (clump A) of some 15 arcmin extent; several  
smaller clumps to the South are connected by tenuous emission.  (Regarding  
the tenuous emission, we note that while structures extending over as much  
as 10~arcmin in a single spectral channel were adequately recovered in the  
WSRT images, there were also indications in the Leiden/Dwingeloo  
single--dish data of more diffuse features which could not be adequately  
imaged by the interferometer. A straightforward attempt was made to correct  
the images for the weak response of the interferometer to diffuse emission  
features, making use of the total flux measured with the Dwingeloo  
telescope.  The integrated \hi flux detected in the reconstructed images  
after primary beam correction varied from less than 1\% to as much as 55\%  
of that detected in the Leiden/Dwingeloo Survey.)  The velocity dispersion  
of the emission from the bright core is less than that from the diffuse  
background. The velocity field of CHVC\,069+04$-$223 shows a systematic  
velocity gradient, oriented along the major axis of clump A from about  
$-$230 to $-$240 \kms\, and extending over some 10 arcmin.  
 
\begin{figure}[]  
\psfig{file=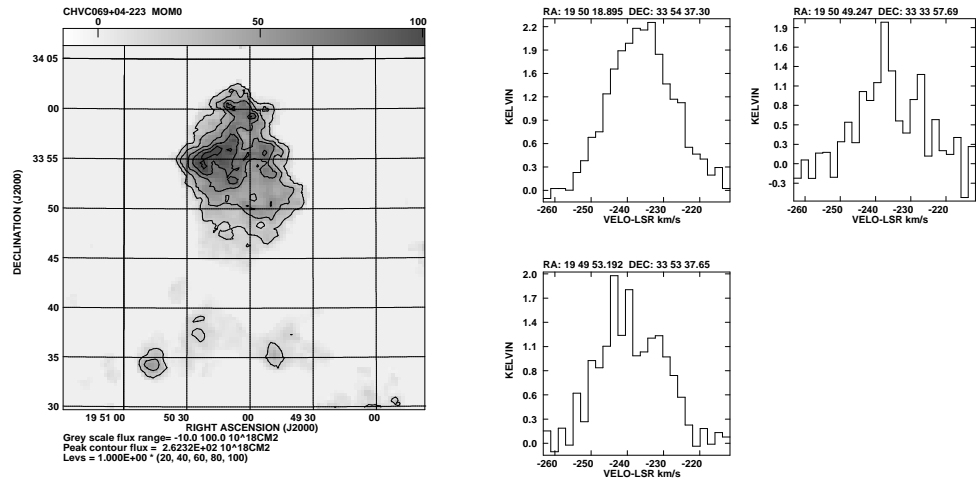,width=13cm}  
\vfill  
\caption{WSRT data for CHVC\,069+04$-$223 at 1~arcmin and 2~km~s$^{-1}$  
  resolution. Left: apparent \NH~ (assuming  
  negligible opacity), with contours at 20, 40, 60, 80, and 100  
  $\times10^{18}$ cm$^{-2}$ and a linear grey--scale extending from $-10$  
  to 100$\times10^{18}$ cm$^{-2}$. Right: brightness  
  temperature spectra at the indicated positions.  
 } \label{fig:h069}  
\end{figure}  
 
%\subsection{CHVC\,115+13$-$275}  
 
    The basic data for CHVC\,115+13$-$275, represented in Fig.~3, similarly  
shows a core/halo morphology.  This source had been completely unresolved  
in the single--dish data, and had shown the broadest linewidth, namely 95  
\kms, of all the CHVCs catalogued by Braun and Burton (1999).  The WSRT  
imaging shows that this FWHM value was contributed by several separate  
knots of emission, each with an intrinsically much narrower velocity FWHM,  
amounting to about 10 \kms.  The collection of substructures, each between  
1 and 10 arcmin in size, is distributed over a region of about 30 arcmin  
extent. Each of the cores has a distinct centroid velocity so that the  
collection spans the velocity interval from $-$300 to $-$220 \kms.  
Detailed  
examination of the WSRT data from the individual cores reveals that several  
of them have significant velocity gradients, oriented preferentially along  
their long axes.  
 
\begin{figure}[]  
\psfig{file=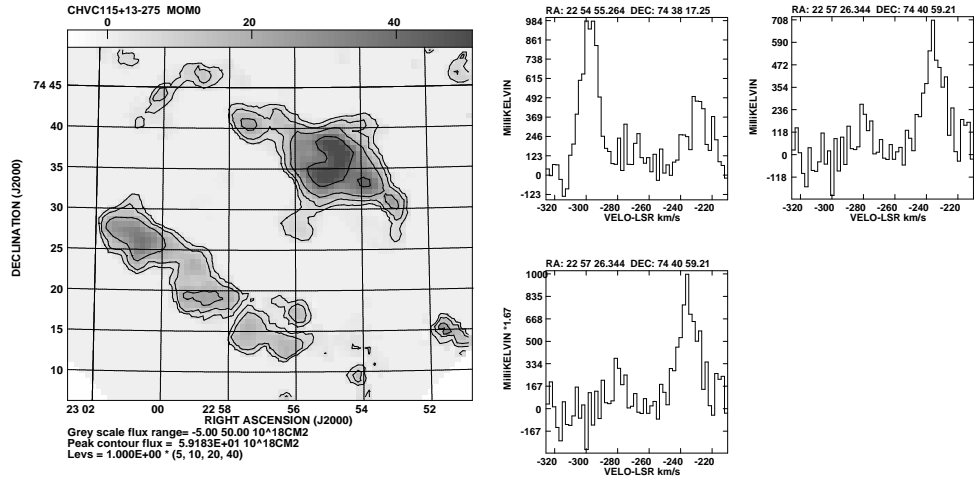,width=13cm}  
\vfill  
\caption{WSRT data for CHVC\,115+13$-$275 at 2~arcmin and 2~km~s$^{-1}$   
resolution. Left: apparent \NH,  with contours at 5, 10, 20, and 40  
$\times10^{18}$ cm$^{-2}$ and a linear grey--scale extending from $-5$ to  
50$\times10^{18}$ cm$^{-2}$. Right: brightness temperature spectra at the  
indicated positions.  The velocity spread of the individually narrow cores  
affords a measure of distance based on dynamical crossing times.  
 } \label{fig:h115}  
\end{figure}  
 
%\subsection{CHVC\,204+30+075}  
 
Figure 4 shows the basic WSRT data for CHVC\,204+30+075.  Several 
relatively  
bright clumps are scattered over an extent of some 35 arcmin.  The spectra  
toward the more compact local maxima have column densities of a few times  
$10^{20}$ cm$^{-2}$, and FWHM widths of less than 15 \kms.  A substantial  
diffuse component was reconstructed for this field, reaching \NH~ values of 
a few times $10^{19}$ cm$^{-2}$ over some 30 arcmin. The large elliptical  
feature in the South--central part of the  
CHVC\,204+30+075 field (clump A) shows a well--defined velocity gradient  
running from about 55 to 80 \kms\, over some 12 arcmin.  Similarly, the  
elliptical feature (clump B) in the North--East region has a velocity  
gradient running from 45 to 75 \kms\, over 20 arcmin.  
 
\begin{figure}[]  
\psfig{file=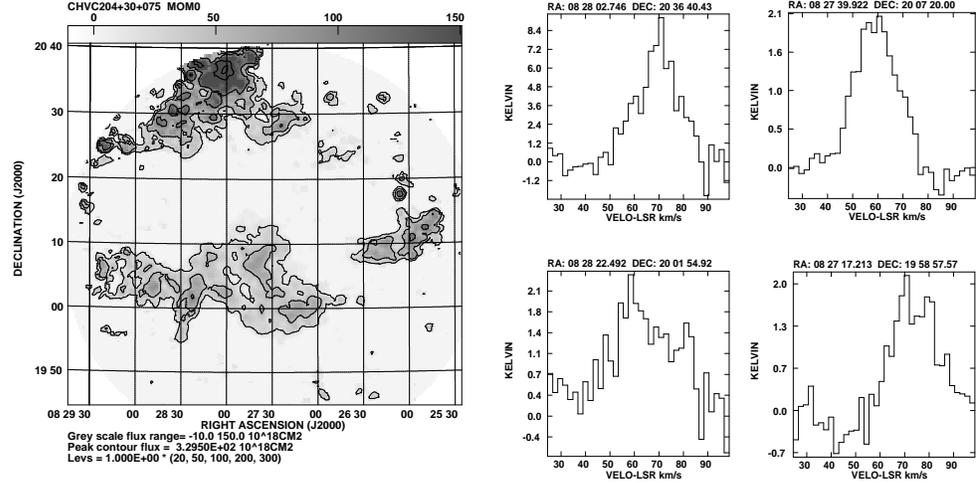,width=13cm}  
\vfill  
\caption{WSRT data for CHVC\,204+30+075 at 1~arcmin and 2~km~s$^{-1}$  
resolution. Left: apparent \NH, with contours at 20, 50, 100, 200, and 300  
$\times10^{18}$ cm$^{-2}$ and a linear grey--scale extending from $-10$  
to 150$\times10^{18}$ cm$^{-2}$. Right: brightness  
temperature spectra at the indicated positions.  
 } \label{fig:h204}  
\end{figure}  
 
\section{Indications of Rotational Support}  
 
     The velocity gradients mentioned above as shown by some of the CHVC 
cores constitute a noteworthy morphological characteristic of  
these objects.  The gradients are oriented along the major axis of roughly  
elliptical \NH\, distributions, and vary in amplitude from 0.5 to 2 \kms  
arcmin$^{-1}$.  Examples of the signature of rotation which are  
particularly well resolved are clump A of CHVC\,069+04$-$223 and clumps A  
and B of CHVC\,204+30+075. These are reminiscent in form and amplitude of  
the ``spider'' diagrams portraying the \hi kinematics of some dwarf  
galaxies.  Consequently we carried out standard tilted--ring fits to assess  
the extent to which the CHVC kinematics could be modeled by rotation in a  
flattened disk system.  
 
     Figure 5 shows that robust solutions for circular rotation pertain for  
the three cases.  The fits indicate rotation velocity rising  
slowly and continuously with radius, to some 15 \kms\, in the case of  
CHVC\,069+04$-$223A, and flattening out to values of 15 and 20 \kms\, for  
CHVC\,204+30+075A and B, respectively.  An estimate of the total mass  
supporting this rotation follows from $M_{\rm dyn}=Rv^2/{\rm G}=2.3 \times  
10^5R_{\rm kpc}v_{\rm km/s}^2$.  At an assumed distance of 0.7 Mpc, these 
cores have $M_{\rm dyn}=10^{7.1}$, $10^{6.5}$, and $10^{6.9}$  
M$_\odot$, respectively.  The mass of the gas, assuming a  
40\% contribution by helium, follows from $M_{\rm gas}=1.4  
M_{\rm HI}=3.2 \times 10^5 S D_{\rm Mpc}^2$, where $S$ is the integrated  
\hi flux in units of Jy\,\kms.  For these three cores, $M_{\rm  
gas}=10^{7.1}$, $10^{6.5}$, and $10^{6.9}$, values which, compared to the 
dynamical masses, correspond to dark--to--visible mass ratios of 10, 36, 
and 29.

\begin{figure}[]  
\vspace{-3.5cm}  
\psfig{file=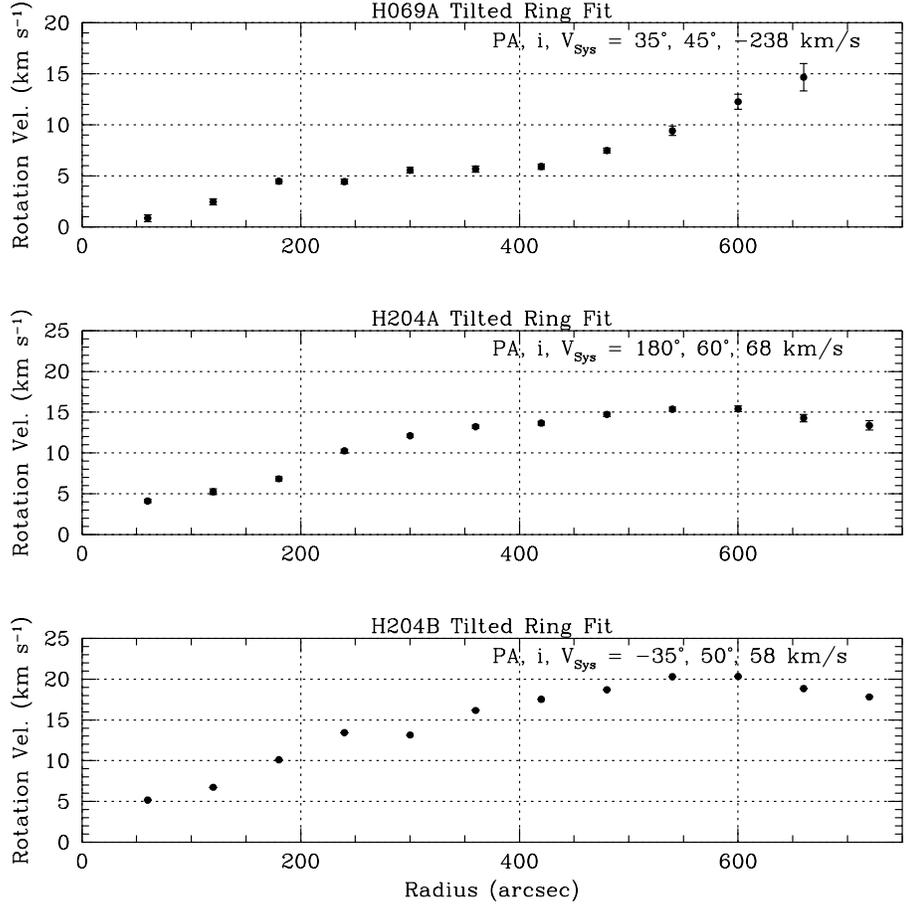,width=13cm}  
\vspace{-0.3cm}  
\vfill   
\caption{Derived rotation velocities in three elliptical cores,  
  CHVC\,069+04$-$223A and CHVC\,204+30+075A and B. The best--fitting  
 position angle, inclination, and systemic velocity are indicated in each 
panel. These were subsequently held fixed in deriving the rotation velocity 
as function of radius.  These CHVC cores are evidently rotationally 
supported.  The dynamical masses derived for the cores assuming a distance 
of 0.7 Mpc are $10^{8.1}$, $10^{8.1}$, and $10^{8.3}$ M$_\odot$; the 
dark--to--visible mass ratios are 10, 36, and 29, respectively.  
 } \label{fig:tilfit}  
\vspace{-2cm}
\end{figure}  
 
\section{Considerations of CHVC Stability}  
 
The very high total linewidth of CHVC\,115+13$-$275 was found to arise
from the range of line--of--sight velocities contributed by the
distinct individual clumps which make up this source.  It is unlikely
that this is a chance superposition of unrelated components, in view of
the spatial and kinematic isolation of this CHVC. Braun and Burton
(2000) consider the stability of these collection of clumps under
several circumstances.  If the collection of clumps were located at a
distance of 5 kpc, the collection would have a diameter of 44 pc and
would double in size on the implausibly short dynamical timescale of
only $5 \times 10^5$ years.  On the other hand, if this source were
located at a distance of 0.7 Mpc and were self--gravitating, then the
angular radius of 15 arcmin and the velocity half--width of 35 \kms\,
could be used to calculate a dynamical mass of $10^{8.93}$ M$_\odot$.
The corresponding mass of gas at this distance, assuming 40\% to be
helium, is $10^{7.22}$ M$_\odot$, and the dark--to--visible mass ratio
is 51 and scales as 1/D.  Consistent distance and dark--to--visible mass
ratios are found for other CHVC objects.
 
\section{Indications of Cold, Opaque Cores}  
 
%\subsection{CHVC\,125+41$-$207}  
 
The object CHVC\,125+41$-$207 is particularly  
interesting.  Of the 65 CHVCs catalogued by Braun and Burton (1999), this  
one showed the narrowest FWHM, amounting to 5.9 \kms\, in the single--dish  
data. The WSRT image shown in Fig.~6 could be displayed at higher angular  
resolution because of the intense brightness of this source.  Like the  
other CHVCs imaged, this one also shows a complex morphology with several  
compact cores.  The spectrum toward the brightest of these cores   
is remarkable in having a linewidth which is completely unresolved  
with the effective resolution of 2.47 \kms.  The velocity channels adjacent  
to the line peak have intensities down to about 20\% of the maximum value.  
Such a width is one of the narrowest ever measured in \hi emission.  
 
The narrow linewidth provides an opportunity (rare in \hi work, where  
linewidths are commonly dominated by line blending and mass motions) to  
derive a kinetic temperature.  An upper limit to the intrinsic FWHM of 2  
\kms\, corresponds to an upper limit to the kinetic temperature of 85 K.   
This limit assumes additional importance because the observed brightness  
temperature of this core is 75 K.  Thus the temperature is tightly  
constrained, and a lower limit to the opacity follows from $T_{\rm 
B}=T_{\rm s}(1-e^{-1})$, if $T_{\rm s}=T_{\rm k}$, yielding $\tau \geq 2$.  
In addition to constraining the temperature, this observation also 
constrains any broadening which might be due to turbulence to be less than 
1 \kms.  
 
The narrow linewidth observed for CHVC\,125+41$-$207 provides an  
opportunity to estimate its distance. Wolfire et al. (1995a, 1995b) show  
that a cold, stable phase of \hi is expected if a sufficient column of  
shielding gas is present and if the thermal pressure is sufficiently high.   
Calculations of equilibrium \hi conditions with the Local Group  
environment have kindly been made available to us by Wolfire, Sternberg,  
Hollenbach, and McKee, for two bracketing values of the shielding column  
density, namely 1 and $10\times10^{20}$ cm$^{-2}$.  Figure 7 shows that the  
corresponding equilibrium volume densities for the observed value of  
$T_{\rm k}=85$ K are 3.5 and 0.65 cm$^{-3}$, respectively.  The distance to  
CHVC\,125+41$-$207 follows from the volume densities thus indicated, and  
from the measured column depths and the angular size of the source,  
according to $D=N_{\rm H}/(n_{\rm H}\theta)$.  Consideration in this way of  
several compact cores in this source allow a distance determination  
in the range 0.5 to 1 Mpc.

\begin{figure}[t]  
\psfig{file=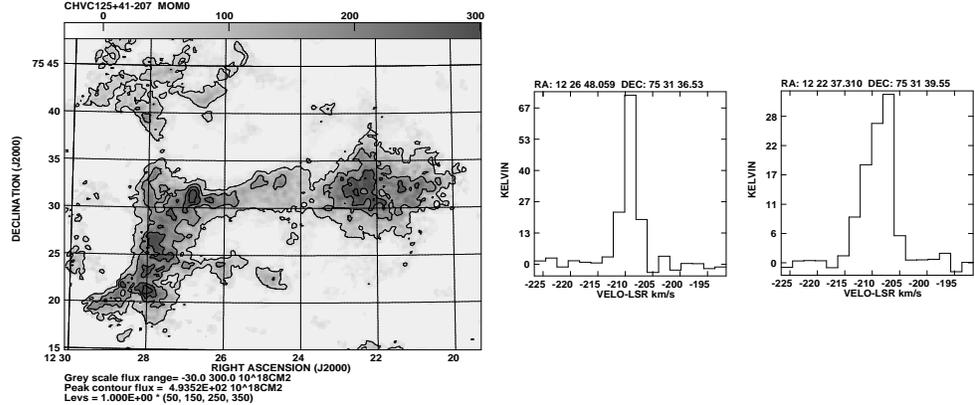,width=13cm}  
\vfill  
\caption{WSRT data for CHVC\,125+41$-$207 at 28~arcsec and 2~km~s$^{-1}$  
resolution. Left: apparent \NH, with contours at 50, 150, 250, and 350  
$\times10^{18}~$cm$^{-2}$ and a linear grey--scale extending from $-30$  
to 300$\times10^{18}~$cm$^{-2}$. Right: brightness  
temperature spectra at the indicated positions. The spectrum shown on the 
left in this panel is one of the narrowest \hi emission lines ever 
observed.  The extremely narrow linewidths robustly constrain the kinetic 
temperatures as well as the turbulence.  
 } \label{fig:h125}  
\end{figure}  
 
\begin{figure}[t]  
\vspace{-3.5cm}  
\psfig{file=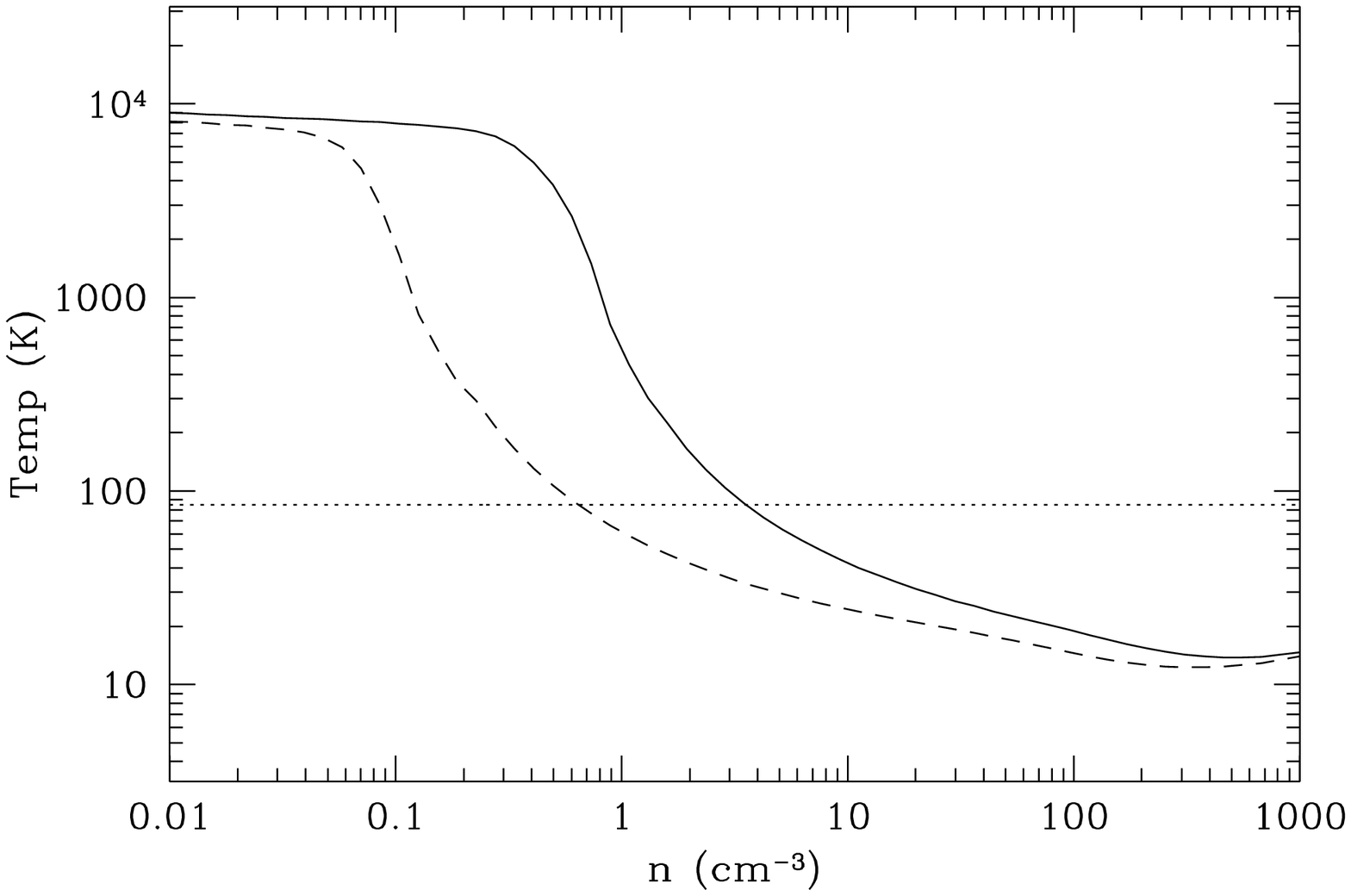,width=13cm}  
\vspace{-1cm}  
\vfill  
\caption{ Equilibrium temperature curves for \hi in an intergalactic  
radiation field at a metallicity of 0.1 solar and a dust--to--gas mass  
ratio of 0.1 times the value representative of the solar neighborhood, and  
for two values of the neutral shielding column density, namely  
$10^{19}$~cm$^{-2}$, indicated by the solid line, and  
$\times10^{20}$~cm$^{-2}$, indicated by the dashed line. The 85~K kinetic 
temperature of the opaque cores in CHVC\,125+41$-$207 is indicated by the 
dotted line.  
 } \label{fig:phased}  
\vspace{-0.5cm}
\end{figure}

\section{Summary:  Parameters of Several of the Individual \\ 
Objects in the CHVC Ensemble}  
 
The compact high--velocity clouds which we have imaged using the Westerbork  
array share several morphological characteristics.  In each case we  
detected a number of compact cores which range in size from a few arcmin to  
about 15 arcmin, with peak column densities in the range $10^{19}$ to  
$10^{20}$ cm$^{-2}$, and FWHM linewidths of about 2 \kms.  Each  
core is characterized by its own line--of--sight velocity; it is the range  
in these core velocities, not the much narrower intrinsic linewidths, which  
determined the range of total widths in the single--dish catalog of CHVCs.  
The modest linewidths seen in all of the cores indicates that the \hi in  
these structures is in the form of the Cool Neutral Medium (CNM) with  
typical equilibrium temperatures in the range 50 to 200 K.  
 
In all of the cases we have imaged, as well as in the two cases imaged  
earlier by Wakker and Schwarz (1991), the CHVC cores are embedded in an  
extended halo of \hi emission.  The halos have typical \hi extents of about  
$1^\circ$, column densities within the inner 30 arcmin between $2 \times  
10^{18}$ and $2 \times 10^{19}$ cm$^{-2}$, and velocity dispersions 
consistent with the 8000 K equilibrium temperature of the Warm Neutral 
Medium (WNM). The cores account for some 40\% of the \hi flux, while 
covering some 15\% of the area. Thus the data at hand suggest that a 
two--phase structure characterizes the CHVC morphology, with cold, opaque 
cores of CNM shielded by a halo of the WNM.  
 
The compact clouds provide several indications of the  
crucial distance parameter. The velocity dispersion of the entire   
ensemble is minimized in a reference frame consistent with the Local Group  
Standard of Rest.  Three additional measures regarding distance are given  
by the imaging of individual CHVCs. $(i)$ Arguments based on the dynamical  
stability of CHVC\,115+13$-$275 support a distance of about 0.7 Mpc. $(ii)$  
The exceptionally narrow linewidths observed in the principal cores of  
CHVC\,125+41$-$207 lead to a measure of the volume density and that, with  
the measured angular size and column depth, yields a distance in the range  
0.5 to 1 Mpc.  $(iii)$ The kinematic gradients detected along  
the long axis of some of the CHVC cores resemble in form and amplitude the  
velocity fields of dwarf galaxies.  The solutions for dynamical mass based  
on this comparison indicate a distance of about 0.7 Mpc. The ratio of 
dark--to--visible mass of the objects is in the range 10 to 50.  

In agreement with the conclusions reached for the ensemble,  
the distances and physical properties of the individual CHVCs are  
suggestive of a population which has as yet had little  
interaction  with the more massive Local Group members. At   
distances of about 0.5 to 1~Mpc these objects would have sizes of about 
15~kpc, gas masses of about 10$^7$~M$_\odot$, and total masses of a few 
times $10^8$~M$_\odot$. We tentatively estimate the total CHVC ensemble to 
harbor some 200 objects; in that case, the total gas mass involved would be 
a few times 10$^9$M$_\odot$.  In view of the net in--fall motion observed 
for the ensemble, a source of dark matter and low--metallicity gas is 
indicated for the continuing  growth and evolution of the major Local Group 
galaxies. The CHVCs may be the missing Local Group satellites predicted by 
hierarchical growth scenarios (e.g. Klypin et al. 1999).  
 
These CHVC parameter values correspond to those of (sub--)dwarf galaxies.   
Indeed, it would be difficult to distinguish these CHVC properties from  
those derived from the \hi signature of a typical dwarf galaxy.  If, on  
the other hand, the objects were produced relatively locally by an  
energetic mechanism responsible for a galactic fountain, then their \hi  
properties would be expected to show large linewidths, motions not  
ordered by rotation, and a structural form other than a core/halo one.  The  
distinction between CHVCs and dwarf galaxies with very weak star formation  
remains to be made and is an important challenge. The CHVCs may represent  
still--pristine examples of collapsed objects, with only a small amount of  
internal star formation and enrichment. As such, they should provide  
insight into the process of galaxy and structure formation.  

The WSRT imaging is discussed in more detail by Braun and Burton (2000).  
 
\acknowledgments  
We are grateful to M.G. Wolfire, A. Sternberg, D. Hollenbach, and C.F. 
McKee for providing the equilibrium temperature curves shown in Fig.~7. 
The Westerbork Synthesis 
Radio Telescope is operated by the Netherlands Foundation for Research in 
Astronomy, under contract with the Netherlands Organization for Scientific 
Research.

\end{document}